\documentclass[twoside,12pt]{article}  
\usepackage{CJK}   
\usepackage{indentfirst}  
\usepackage{bm}    
\usepackage{graphicx}  
\usepackage{float}

\renewcommand{\baselinestretch}{1.5}     

\begin{document}    

\begin{CJK*}{GBK}{song}  


=============================================

\begin{center}
\LARGE\bf On the vibron dressing in the $\alpha$--helicoidal macromolecular
chains$^{*}$   
\end{center}

\footnotetext{\hspace*{-.45cm}\footnotesize $^*$This work was supported by the Ministry of Education and Science of Republic of Serbia under contract numbers III--45005 and III--45010, and by the project within the Cooperation Agreement between the JINR, Dubna, Russian Federation and Ministry of Education and Science of Republic of Serbia. The study was supported by The Ministry of education and science of Russian
Federation, project 14.B37.21.0774 and 14.B37.21.1301.  The authors are grateful to I. Bondarev, for valuable comments and discussion on possibility of coexistence two vibron states, mobile and immobile ones. }
\footnotetext{\hspace*{-.45cm}\footnotesize  $^\dag$Corresponding author. E-mail: cevizd@vinca.rs;\quad $^\ddag$E-mail:reshet@ispms.tsc.ru}

\begin{center}
\rm D. \v Cevizovi\'c$^{\rm a),\rm b) \dagger}$, \ \ S. Galovi\'c$^{\rm a),\rm b)}$, \ \ A. Reshetnyak$^{\rm c) \ddagger}$ \ and \ Z. Ivi\'c$^{\rm a)}$
\end{center}

\begin{center}
\begin{footnotesize} \sl
$^{\rm a)}${University of Belgrade, "Vin\v ca" Institute of Nuclear Sciences Laboratory for theoretical and condensed matter physics, P.O. Box 522, 11001 Belgrade, Serbia}\\   
$^{\rm b)}${Joint Institute for Nuclear Research Bogoliubov Laboratory for Theoretical Physics, Dubna, Russia}\\   
$^{\rm c)}${Laboratory of Computer-Aided Design of Materials, Institute of Strength Physics and Materials Science, Tomsk, 634021 , Russia} \\   
\end{footnotesize}
\end{center}

\begin{center}
\end{center}

\vspace*{2mm}

\begin{center}
\begin{minipage}{15.5cm}
\parindent 20pt\footnotesize
{We present a study of the physical properties of the vibrational excitation in $\alpha$--helicoidal macromolecular chains, caused by the interaction with acoustical and optical phonon modes. The influence of the temperature and the basic system parameters on the vibron dressing has been analyzed by employing the simple mean--field approach based on the variational extension of the Lang--Firsov unitary transformation. Applied approach predicts a region in system parameter space where one takes place an abrupt transition from partially dressed (light and mobile) to fully dressed (immobile) vibron states. We found that the boundary of this region depends on system temperature and type of bond among structural elements in the macromolecular chain.}
\end{minipage}
\end{center}

\begin{center}
\begin{minipage}{15.5cm}
\begin{minipage}[t]{2.3cm}{\bf Keywords:}\end{minipage}
\begin{minipage}[t]{13.1cm}
 Vibron, Small-polaron, $\alpha$--helix, partial dressing 
\end{minipage}\par\vglue8pt
{\bf PACS: 05.30.-d, 31.15.xt, 63.20.kk, 66.30.hp, 71.38.Ht}
\end{minipage}
\end{center}

\section{Introduction}  
It is believed that the hydrolysis of the adenosine triphosphate (ATP) is the universal mechanism providing the energy for diverse biological processes such as photochemical reactions, cross--membrane ion transfer and signal transduction, muscle contraction, cellular mobility and transports to mention just a few \cite{DavydovBQM}. All these processes involve protein molecules as mediator of the long distance energy transfer. However, it is yet not clearly understood how this energy can be transported along the polypeptide chain at long distances, without being dissipated or dispersed.

An early explanation of this problem, based on quantum mechanical model, has been proposed in the mid--seventies by Davydov and collaborators \cite{DavS1,DavS2,DavS3}. The essence of their theory is the assumption that the energy released by the ATP hydrolysis can be captured by the protein molecules where it excites the high--frequency amide--I vibrational mode ($\mathrm{C}=\mathrm{O}$ stretching quanta or vibrons) of a peptide group. Due to the dipole--dipole coupling this energy delocalizes between the adjacent peptide groups giving rise to vibron band states. Such collective excitations may coherently propagate along the polypeptide chain participating in such a way in the energy transfer. Nevertheless, as it was shown by Davydov and collaborators \cite{DavydovBQM,DavS1,DavS2,DavS3}, their lifetime is to short which rules out the long range energy transfer via band--like mechanism. Instead, they suggested that the energy losses of the vibron may be prevented by its (self--)trapping in the potential well created by the induced local distortion of the molecular crystals. So created complex entity, i.e. the vibron surrounded with the local lattice distortion, may propagate in a soliton form along the chain with minimal energy losses preserving its shape and velocity for a long time.

According to the general theory of self--trapping (ST) phenomena \cite{Rashba,Firsov}, the character of ST states is determined by the mutual ratio of the values of three basic energy parameters of the system. These parameters are the exciton (electron, vibron, etc.) bandwidth $2|J|$ and characteristic phonon energy $\hbar\omega_C$ (determined by the phonon cut--off frequency $\omega_C$) characterizing respectively, the time scale of motion in excitation and lattice subsystems. The third parameter is lattice deformation energy $E_B$ which measures the strength of vibron--phonon interaction \cite{Rashba,Firsov,EminAP22,HolsteinAP8_1,HolsteinAP8_2,LF}. Two limiting cases could be considered as well understood: large-- and small--polaron limit \cite{Rashba,Firsov,EminAP22,HolsteinAP8_1,HolsteinAP8_2,LF}. The first one is reached when the excitation bandwidth and lattice deformation energy both highly exceeds the characteristic phonon energy i.e. in the adiabatic strong coupling regime. In that case phonons are slow with respect to the excitation dynamics and form, essentially classical, large radius quasi--static potential well in which that excitation (vibron) is trapped. Contrary, in the non--adiabatic limit, when the characteristic phonon energy is larger than the vibron bandwidth, the quantum nature of the phonons plays a crucial role, and small polaron (SP) formation takes place. They are characterized by a short--range lattice distortion that follows polaron motion instantaneously.

The vibron bandwidth in hydrogen--bonded macromolecules such as polypeptide chains is about 0.001 eV, while the phonon cut--off frequencies are typically about $10^{13}$ Hz \cite{ScottPR217,ScottPRA26,NevskayaBP15}. As a consequence, basic assumptions for the soliton formation are not fulfilled \cite{BIPRB40}. For that reason, it has been suggested by some authors that vibron self--trapping in hydrogen--bonded macromolecular chains might result in formation of a small--polaron, rather than a soliton \cite{BIPRB40,IvicPD113,AlexanderPRB33,IvicPRB48,IvicJPCM9,TekicPRE60}. The applicability of such concept is restricted to the strong coupling and non--adiabatic limit which imposes certain doubts on the whole SP concept since only the non--adiabatic condition is satisfied, while the problem of the strength of vibron--phonon coupling in these structures is still under the debate.

Usually, the studies devoted to vibron excitation problem in macromolecular chains involve the value of the strength of vibron--phonon coupling constant extracted from an indirect comparison between experimental and theoretical results \cite{AlexanderPRB33}. According to these estimations, the value of this constant in hydrogen bonded macromolecular chains ranges between 30 pN and 60 pN \cite{AlexanderPRB33,CruzeiroJCP123}. However, there are such estimations of that parameter, performed by the \textit{ab initio} calculations for the formamide dimmer, which suggests that its value is significantly lower \cite{Kuprievich,Pierce,Ostergard}. This fact became by the motivation for some authors to define criteria to discriminate between the weak and the strong vibron--phonon coupling limits \cite{ScottPR217,PouthierJCP132}. According to \cite{PouthierJCP132}, the strength of the vibron--phonon coupling in a lattice of $H$--bonded peptide units at room temperatures belong to strong coupling limit, but uncertainty of the value of the coupling constant may leads to completely different conclusions. So, this problem is still unresolved and  requires a further research. Additionally, some recent numeric \cite{HammPRB78} and theoretical \cite{CevizovicPRE84} studies of the ST phenomena in hydrogen--bonded macromolecular chains indicate that the character of ST states can be qualitatively very different depending on the values of system parameters, temperature and, therefore its proper theoretical description requires an approach that goes beyond the conventional strong--coupling SP theories. Especially useful in that sense is the concept of partial dressing, which is applied in a quite wide part of system parameter space. It relies on modified Lang--Firsov (MLF) unitary transformation. Such an approach is in close correspondence with the supplementary variational treatment of the problem by means of the Toyozawa ansatz \cite{ToyozawaPTP26}.

In this paper we study the character of vibron states in $\alpha$--helicoidal macromolecular chains. We supposed that vibron interacts with acoustic and optical phonon modes. The problem of the vibron interaction with optical phonon modes has been intensively considered in many publications in the framework of the Holstein model \cite{ScottPR217,HammPRB73,HammEPJst147}, especially after the indications that in $ACN$ the vibron interaction with optical phonon modes plays a crucial role for the vibron small--polaron formation \cite{BarthesJML41,EdlerPRL88}. On the other side, the problem of the vibron interaction with acoustical phonon modes usually has not been considered. This problem, in our opinion  should not be disregarded since acoustical phonon modes have quite lower energies in comparison with the optical ones, and consequently, can be easily excited in the whole temperature range.

Our study is performed within the  D=1 dimensional model resembling the real $\alpha$--helix structure. We assumed that, due to the dipole--dipole interaction, the vibron excitation could be delocalized either from a some structural element to its first--neighbor ones (from $n$-th to ($n\pm 1$)-th structural element), or from a some structural element to its non--neighboring (from $n$-th to $m\neq(n,n\pm1)$-th) structural elements. This light generalization of simple one-dimensional model enables treatment of the vibron delocalization along the direction defined by the covalent bonds in $\alpha$--helix (corresponding to the vibron delocalization between the first neighboring structural elements), while the vibron delocalization along the hydrogen bonds in $\alpha$--helix macromolecule corresponds to vibron delocalization between each third neighboring structural elements in the present model. We restrict our attention to the single--vibron case.

Our particular interest is to establish conditions when vibron moves in a coherent (via band states) or incoherent (diffusive motion by means of the random hops between neighboring peptide groups) fashion along such structure. Special attention will be  paid to the examination of the phenomenon known as small polaron crossover, which may occur for the certain values of system parameters and in the certain temperature range. This feature is connected with known ambiguities in SP parameters, effective mass, for example, arising as a consequence of the sudden jump in the magnitude of the degree of phonon dressing.

\section{Model Hamiltonian}

The system under the consideration consists of single vibron excited on $n$-th structural element of the macromolecule,  which physical properties are affected by the thermal vibrations of the macromolecular chain. We suppose that vibron excitation can move along the chain from $n$-th to the $(n\pm1)$-th structural element, either from $n$-th to the $(n\pm3)$-th structural element. The first possibility corresponds to vibron delocalization along the covalent bonds, while the second possibility corresponds to vibron delocalization along the hydrogen bonds in 3D $\alpha$--helix structure. Due to the fact that the energy of the dipole--dipole interaction between second--neighboring peptide groups is smaller than the energy of the dipole--dipole interaction between the first--neighboring ones \cite{ScottPRA26}, the vibron delocalization from $n$-th to ($n\pm2$)-th structural element is not considered. Corresponding Hamiltonian of the vibration excitation and phonon subsystem can be written in the following form \cite{HolsteinAP8_1}:
\begin{eqnarray}\label{PocHam}
 H & = & \Delta\sum_n{A^{\dagger}_nA_n}-\sum_n{J_gA^{\dagger}_n(A_{n+g}+A_{n-g})}+
 \sum_q{\hbar\omega_qB^{\dagger}_qB_q} \nonumber \\
 && +
 \frac{1}{\sqrt{N}}\sum_q\sum_n{F_q\mathrm{e}^{iqnh}A^{\dagger}_nA_n(B_q+B^{\dagger}_{-q})}.
\end{eqnarray}
In the Hamiltonian above the quantity $\Delta$ is the vibron excitation energy, $A^{\dagger}_n(A_n)$ describes the presence (absence) of the vibron quanta on the structural element, which is positioned on $n$-th lattice site, $B^{\dagger}_q(B_q)$ creates (annihilates) phonon quanta, and $\omega_q$ is the phonon frequency. The inter--site overlap integral $J_g$ characterize the vibron transfer between macromolecular chain structural elements: $g=1$ in the case of the vibron transfer along the covalent bonds, and $g=3$ in the case of the vibron transfer along the hydrogen bond. Delocalization along the covalent and hydrogen bonds will be examined separately, and therefore the summation over the index $g$ was not included in (\ref{PocHam}). This refers to  two practically independent models describing the vibron delocalization along particular bonds. In such a way, the mutual influence of the vibron delocalization along covalent bonds on vibron delocalization along hydrogen bonds (and vice versa) has been disregarded. Widely accepted data for values of the hopping constant $J_3$ is $J_3=J_{hb}=7.8\; \mathrm{cm}^{-1}$, while for the hopping constant between different spines of hydrogen--bonded peptide units is $J_1=J_{cov}=-12.4\;\mathrm{cm}^{-1}$ \cite{ScottPR217,ScottPRA26,NevskayaBP15}. The parameter $h$ is a distance between two neighboring structural elements along the axis of the macromolecular chain. Finally, $F_q=F^*_{-q}$ is the vibron--phonon coupling parameter, which is a function of the coupling constants $\chi_g$. Coupling constant accounts for the modulation of the vibron frequency placed on the $n$-th structural element due to the external motion of the $(n+g)$-th structural element. The value of the coupling constant that determines the vibron--phonon coupling in the direction defined by hydrogen bonds ($\chi_{hb}$) in $\alpha$--helix structure is known \cite{AlexanderPRB33,CruzeiroJCP123}. However, the influence of the motion of the $(n\pm 1)$-th and $(n\pm 2)$-th structural elements on the amide--I vibration on the $n$-th structural element is still unknown. According some authors \cite{PouthierJCP123}, numerical values of these parameters satisfy following conditions: $\chi_{hb}>\chi_1>>\chi_2$. In this paper we assume that $\chi_1=0$, $\chi_2=0$ and $\chi_3=\chi_{hb}$. Consequently, in the case of the vibron interacting with acoustic phonon modes, $F_q$ becomes as, $F_q=2i\chi_{hb}\sqrt{\frac{\hbar}{2M\omega_q}}\sin{(3qh)}$ \cite{AlexanderPRB33,IvicPRB48}, while in the case of the vibron interacting with optical phonon modes this parameter becomes: $F_q=\chi_{hb}\sqrt{\frac{\hbar}{2M\omega_q}}$. Phonon dispersion law is determined by the geometric structure of the macromolecular chain, mass of the structural elements from which the chain is built, as well as the values of the elasticity coefficients that characterize the stiffness of the covalent and hydrogen bonds in the chain. Due to the fact that in the case of $\alpha$--helix secondary structure three intermolecular interactions is the minimal number of bonds required to have a stabilized structure \cite{ChristiansenPRE56}, we assumed that each structural element is connected with its first--neighbors, second--neighbor and third--neighbors. The elasticity coefficient $K_1$ of the covalent bonds, that connects nearest--neighbor peptide units, varies between 45 N/m and 75 N/m \cite{HennigPRB65}, while the elasticity coefficient $K_3$ of the hydrogen bonds ranges from 13 N/m to 20 N/m \cite{ScottPR217,IvicJPCM9,PouthierJCP123,HennigPRB65}. In contrast to the analysis carried out in the \cite{PouthierJCP123}, we take into account non--zero value of the elasticity coefficient $K_2$ of the second--neighbor bonds which presence can significantly affect on the nature of the phonon spectra.  We vary this value from interval $K_2\in[\mathrm{min}K_2,\mathrm{max}K_2]$, where $\mathrm{min}K_2<<K_3$ and $K_1>\mathrm{max}K_2>>K_3$. Consequently, dispersion law for acoustic phonon modes have following form: $$\omega^2_q=\frac{4K_1}{M}\sin^2\left(\frac{qh}{2}\right)+\frac{4K_2}{M}\sin^2\left(qh\right)+\frac{4K_3}{M}\sin^2\left(\frac{3qh}{2}\right).$$ In the case of optical phonon case we used the dispersioneless approximation: $\omega_q=\omega_C$, where phonon cut--off frequency is $\omega_C=2\sqrt{\frac{K_1+4K_2+9K_3}{M}}$ \cite{DavydovTTT}. Here, $M$ is the mass of the peptide unit. For the mass parameter we adopted the following numerical value: $M=5.7\cdot 10^{-25}$ kg \cite{ScottPR217}.

In order to examine under which conditions dressed vibron excitations represent dynamically stable eigenstates of the system, we cross to small--polaron picture using MLF unitary transformation operator \cite{CevizovicPRE84} $$U=\exp\left\{{-\frac{1}{\sqrt{N}}\sum_q\sum_n{f_q\mathrm{e}^{-iqnh}A^{\dagger}_nA_n(B_{-q}-B^{\dagger}_q)}}\right\}$$ and we rewrite (\ref{PocHam}) in terms of new operators $a_n=UA_nU^{\dagger}$ ($a^{\dagger}_n=UA^{\dagger}_nU^{\dagger}$) and $b_q=UB_qU^{\dagger}$ ($b^{\dagger}_q=UB^{\dagger}_qU^{\dagger}$), representing the dressed vibrons and new phonons in the lattice with shifted equilibrium positions of molecular groups.

Here $f_q=\delta\frac{F^*_q}{\hbar\omega_q}$, $0<\delta<1$ is variational parameter, which characterizes the degree in which the vibron distorts the lattice and the lattice feedback on the vibron, i.e. vibron dressing.

On the next stage, we apply a simple mean--field approach, which enables the explicit accounting for the influence of thermal fluctuations on SP properties and, in particular, the degree of dressing. In the present case, the temperature enters in  our model Hamiltonian implicitly through phonon fluctuations around its new equilibrium positions. Thus, the explicit temperature dependence may be introduced here by the appropriate averaging of the transformed Hamiltonian over the phonon subsystem. In particular, we define (in wave number representation) an effective, mean--field Hamiltonian $H_0$ as follows:
\begin{equation}\label{MFH}
H=H_0+H_{int},
\end{equation}
where $H_0=H_v+H_{ph}$, $H_v=\left\langle H-H_{ph}\right\rangle_{ph}$ and $H_{int}=H-H_{ph}-\left\langle H-H_{ph}\right\rangle_{ph}$. The symbol $\left\langle \  \ \right\rangle_{ph}$ denotes the average over the new--phonon ensemble, which is in thermal equilibrium state at temperature $T$. In such a way we obtain
\begin{eqnarray}\label{H0}
H_0 &= & \sum_k{E^g_{SP}(k)a^{\dagger}_ka_k}+\sum_q{\hbar\omega_qb^{\dagger}_qb_q}\,,\\
\label{Hint}
H_{int}&= & \frac{1}{\sqrt{N}}\sum_q\sum_k{(F^*_q-\hbar\omega_qf_q)a^{\dagger}_ka_{k+q}(b_{-q}+b^{\dagger}_q)} -\sum_{k_1}\sum_{k_2}J_ga^{\dagger}_{k_1}a_{k_2}\left\{\mathrm{e}^{ik_2gh}\left(T_{k_1,k_2}(g)\right.
\right.
\\ \nonumber
&& \left.\left. -\mathrm{e}^{-W_g(T)}\delta_{k_1,k_2}\right)+
\mathrm{e}^{-ik_2gh}\left(T_{k_1,k_2}(-g)-\mathrm{e}^{-W_g(T)}\delta_{k_1,k_2}\right)\right\},
\end{eqnarray}
 where $a_k=\frac{1}{\sqrt{N}}\sum_n{\mathrm{e}^{iknh}a_n}$, and operators  $T_{k_1,k_2}(\pm g)$ look as
 \begin{eqnarray}\label{Tk1k2}
T_{k_1,k_2}(\pm g)=\frac{1}{N}\sum_n{\mathrm{e}^{i(k_2-k_1)nh} \theta^{\dagger}_{n\pm g}\theta_n},\texttt{ where } \theta_n=\mathrm{e}^{ \frac{1}{\sqrt{n}} \sum_q{f_q\mathrm{e}^{iqnh}(b_q-b^{\dagger}_{-q})}},
\end{eqnarray}
 while mean values of $T_{k_1,k_2}(\pm g)$ over phonon subsystem calculated as follows
 \begin{eqnarray}\label{avtk1k2}
 \left\langle T_{k_1,k_2}(\pm g)\right\rangle_{ph}=\frac{1}{\sqrt{N}}\sum_n{\mathrm{e}^{i(k_2-k_1)nh}\left\langle\theta^{\dagger}_{n\pm g}\theta_n\right\rangle_{ph}}= \mathrm{e}^{-W_g(T)}\delta_{k_1,k_2}.
 \end{eqnarray}
 The variational energy of the small--polaron band state is given by the formula,
\begin{equation}\label{ESP}
E^g_{SP}(k)=\Delta-\frac{1}{N}\sum_q{\left\{(f_q+f^*_{-q})F_q-\hbar\omega_q\left|f_q\right|^2\right\}}-2J_g\mathrm{e}^{-W_g(T)}\cos(gkh) \end{equation}
 with quantity
\begin{equation}\label{W}
W_g(T)=\frac{1}{N}\sum_q{\left|f_q\right|^2(2\bar{\nu}_q+1)(1-\cos(gqh))}
\end{equation}
denoting the vibron--band narrowing factors, which characterizes the degree of the reduction of the corresponding overlap integrals or equivalently the enhancement of the polaron effective mass. For obvious reasons, they are sometimes called the "`dressing fractions'" or the "`dressing parameters'". At last, $\bar{\nu}_q=1/(\mathrm{e}^{\hbar\omega_q/k_BT}-1)$ denotes the phonon average number.

Partially dressed vibron states represent dynamically stable eigenstates of the system provided that variational parameter corresponds to the minimum of the system free energy. Thus, we search for its optimized value minimizing the system free energy, $\mathcal{F}=-k_BT\ln\mathrm{Tr}{\{\mathrm{e}^{-H/k_BT}\}}$. According to Bogoliubov variational theorem (which states that the function $$\mathcal{F}_B=-k_BT\ln{\mathrm{Tr}\{\mathrm{e}^{-H_0/k_BT}\}}+\langle H_{int}\rangle_{H_0}$$ determines the upper bound of the system free energy: $\mathcal{F}\leq\mathcal{F}_B$), the procedure of the minimization of the $\mathcal{F}$ can be replaced by the procedure of the minimization of the $\mathcal{F}_B$. In above expression, $\langle H_{int}\rangle_{H_0}$ denotes the mean value of the interaction polaron--phonon term, over non--interacting polaron--phonon system $H_0$.

Using vector basis $\{|\psi(k,\{n_q\})\rangle=a^{\dagger}_k |0\rangle_{vib}\otimes \prod_q\frac{(b^{\dagger}_q)^{n_q}}{\sqrt{n_q!}}|0\rangle_{ph}|, \forall k, \forall n_q\}$, one can obtain following form of $\mathcal{F}_B$:
\begin{equation}\label{FgB}
\mathcal{F}^g_B=-k_BT\ln\sum_k{\mathrm{e}^{-E^g_{SP}(k)/k_BT}}-k_BT\sum_q\ln\frac{1}{1-\mathrm{e}^{-\hbar\omega_q/k_BT}}.
\end{equation}
It is useful to have in mind that only first term of $\mathcal{F}^g_B$ depends on variational parameter $\delta$.

\section{Results and discussion}

In order to analyze the character of the dressed vibron states in macromolecular chains, we examine the degree of the vibron--polaron dressing (i.e. its effective mass) and the vibron--polaron ground state energy. Note,  the ground--state vector of the effective Hamiltonian $H_0$ reads $\left|\psi_{GS}\right\rangle=a^{\dagger}_{k_0}\left|0\right\rangle_{vib}\otimes\prod_q\left|0\right\rangle_q$, where $\prod_q\left|0\right\rangle_q$ is the phonon vacuum vector and $k_0=\pi/3h$ ($g=3$), or $k_0=0$ ($g=1$). This case is also the most relevant for the spectroscopy since vibron bandwidths in these media are very narrow and its dispersion in all practical examination of optical spectra can be neglected \cite{EminAP22,ScottPR217,HammPRB78,MahanMPP,Fitchen}. As a consequence the ground--state energy corresponds to the lowest level of SP energy, i.e., $E^g_{GS}=\left\langle\psi_{GS}\left|H_0\right|\psi_{GS}\right\rangle=E^g_{SP}(k_0)$.

Further calculations are done in terms of two independent dimensionless parameters: adiabatic parameter $B=\frac{2\left|J\right|}{\hbar\omega_B}$  determining the character of the lattice deformation engaged in the polaron formation and coupling constant $S=\frac{E_B}{\hbar\omega_B}$ (where $E_B=\frac{1}{N}\sum_q{\frac{\left|F_q\right|^2}{\hbar\omega_q}}$ is the lattice deformation energy), which determines the polaron spatial size and the degree of dressing. This parameter, which has originally introduced by Holstein \cite{HolsteinAP8_1,HolsteinAP8_2} has special importance in the strong--coupling non--adiabatic regime ($S>>1$ and $B<<1$) where it characterizes the SP narrowing of quasi particle band which decrease exponentially as $\mathrm{e}^{-S}$.

The expressions for the dressing fractions and normalized ground state energy are calculated for vibron which delocalizes along the covalent bonds, and for the vibron which delocalizes along the hydrogen bonds. We take into account the vibron which interacts with acoustical and optical phonon modes separately. The variational parameter $\delta$ is calculated by the minimization of the Bogoliubov upper bound of the system free energy: $\frac{\partial \mathcal{F}_B}{\partial \delta}=0$ and $\frac{\partial^2\mathcal{F}_B}{\partial\delta^2}>0$.

A normalized ground state energy (normalized on the characteristic phonon energy and measured from the vibron excitation energy level: $\mathcal{E}^{g}_{GS}=\frac{E^{g}_{GS}-\Delta}{\hbar\omega_C}$) takes the value,
\begin{equation}\label{EGS}
\mathcal{E}^{g}_{GS}(\tau)=-S(2-\delta)\delta-B\mathrm{e}^{-W_g(\tau)}
\end{equation}

The first term in (\ref{EGS}) corresponds to quasiparticle binding energy, while the second one corresponds to the width of the quasiparticle energy band. In the weak coupling limit or for intermediate coupling strength, there exists quasiparticle energy band of certain width. Consequently, in that case quasiparticle is delocalized (it can move along the chain on the coherent band--like fashion). On the other side, in the strong coupling limit (i.e. for the full dressed vibron case when $\delta=1$) energy band is quite narrow, and consequently $\mathcal{E}_{GS}$ is proportional to quasiparticle binding energy \cite{LF,BIPRB40}. In that case quasiparticle is almost localized on the particular macromolecular structural element (i.e. on the particular peptide unit), and it can move by random jumps between neighboring peptide units. Functional dependence of normalized ground state energy on coupling constant becomes $\mathcal{E}_{GS}(S)\cong -S$. These simple conclusions may be very useful in analyzing of the obtained results and explanation of the nature of the dressed vibron states.

The expression for vibron dressing fraction, in the case of the vibron which delocalizes along the covalent bond reads
\begin{equation}\label{WCovAK}
W(\tau)=\delta^2S\frac{I_W(\tau)}{I_{Eb}},
\end{equation}
when vibron interacts with acoustical phonon modes, while the quantity
\begin{equation}\label{WCovOPT}
W(\tau)=\delta^2S\coth\left(\frac{1}{2\tau}\right)
\end{equation}
is vibron dressing fraction when vibron interacts with optical phonon modes. In (\ref{WCovAK}) integrals $I_W(\tau)=\frac{1}{\pi}\int_0^{\pi}{\frac{1}{\bar{\omega}^3_x}\coth\left({\frac{\bar{\omega}_x}{2\tau}}\right)\sin^23x(1-\cos x)}dx$ and $I_{Eb}=\frac{1}{\pi}\int^{\pi}_0{\frac{\sin^23x}{\bar{\omega}^2_x}dx}$ are obtained by the virtue of the standard  rule, $\frac{1}{N}\sum_q...\rightarrow\frac{h}{2\pi}\int^ {\pi/h}_{-\pi/h}...dq$.

In the case when vibron delocalizes along the hydrogen bonds, vibron dressing fraction becomes
\begin{equation}\label{WHAK}
W(\tau)=\delta^2S\frac{\bar{I}_W(\tau)}{I_{Eb}},
\end{equation}
when vibron interacts with acoustical phonon modes, and in the case when vibron interacts with optical phonon modes it takes the form
\begin{equation}\label{WHOPT}
W(\tau)=\delta^2S\coth\left(\frac{1}{2\tau}\right).
\end{equation}
The integral in equation (\ref{WHAK}) is $\bar{I}_W(\tau)=\frac{1}{\pi}\int_0^{\pi}{\frac{1}{\bar{\omega}^3_x}\coth\left({\frac{\bar{\omega}_x}{2\tau}}\right)\sin^23x(1-\cos 3x)}dx$. The system normalized temperature is $\tau=\frac{k_BT}{\hbar\omega_C}$, $\bar{\omega}_x=\frac{\omega_x}{\omega_C}$ and $x=qh$.

Obtained results are illustrated on figures Fig.1-6. On figures Fig.1-4, the dressing fractions and ground state energy are presented by the set of "`adiabatic"' curves $W(S)$ and $\mathcal{E}_{GS}(S)$ (each curve corresponds to a particular fixed value of the adiabatic parameter $B$) for various values of $B$, system temperature $T$, for both types of the bonds between peptide units, and for both cases of the vibron--phonon interaction (interaction with acoustical and optical phonon modes). The adiabatic curves $\mathcal{E}_{EG}(S)$ obtained by the standard LF theory are presented by the thin curves, while the ones, obtained by the variational approach are represented by the thick curves.

\begin{figure}[H]
\begin{center}
  \includegraphics[height=.25\textheight]{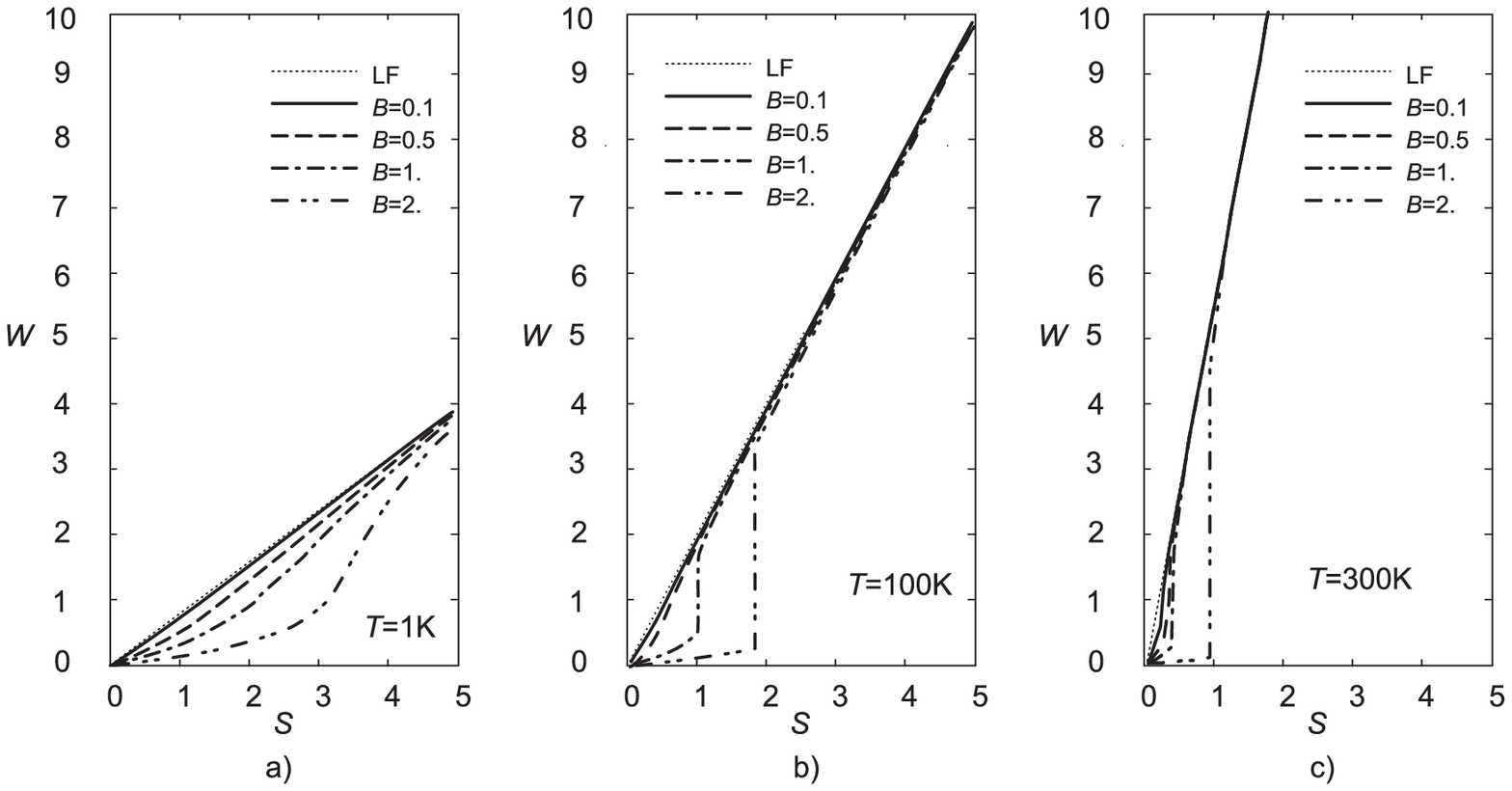}
  \includegraphics[height=.25\textheight]{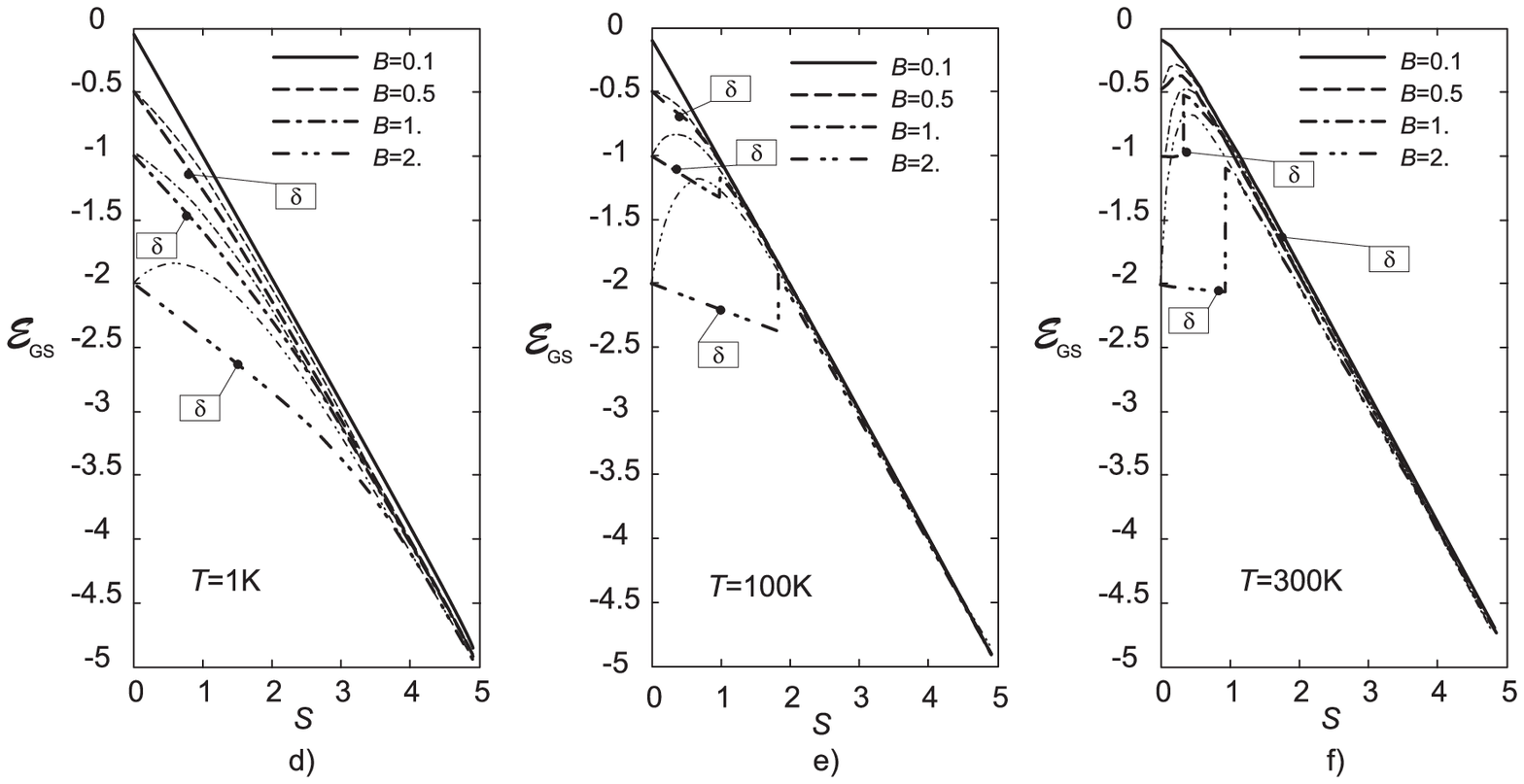}
  \caption{Dependence of the vibron dressing fraction $W$ (first row), and the vibron ground state energy $\mathcal{E}_{GS}$ (second row), on the coupling constant $S$, for various $B$ and $T$. The case of the vibron which moves along the covalent bonds and interacts with acoustical phonon modes.}\label{fig_1}
\end{center}
\end{figure}

\begin{figure}[H]
\begin{center}
  \includegraphics[height=.25\textheight]{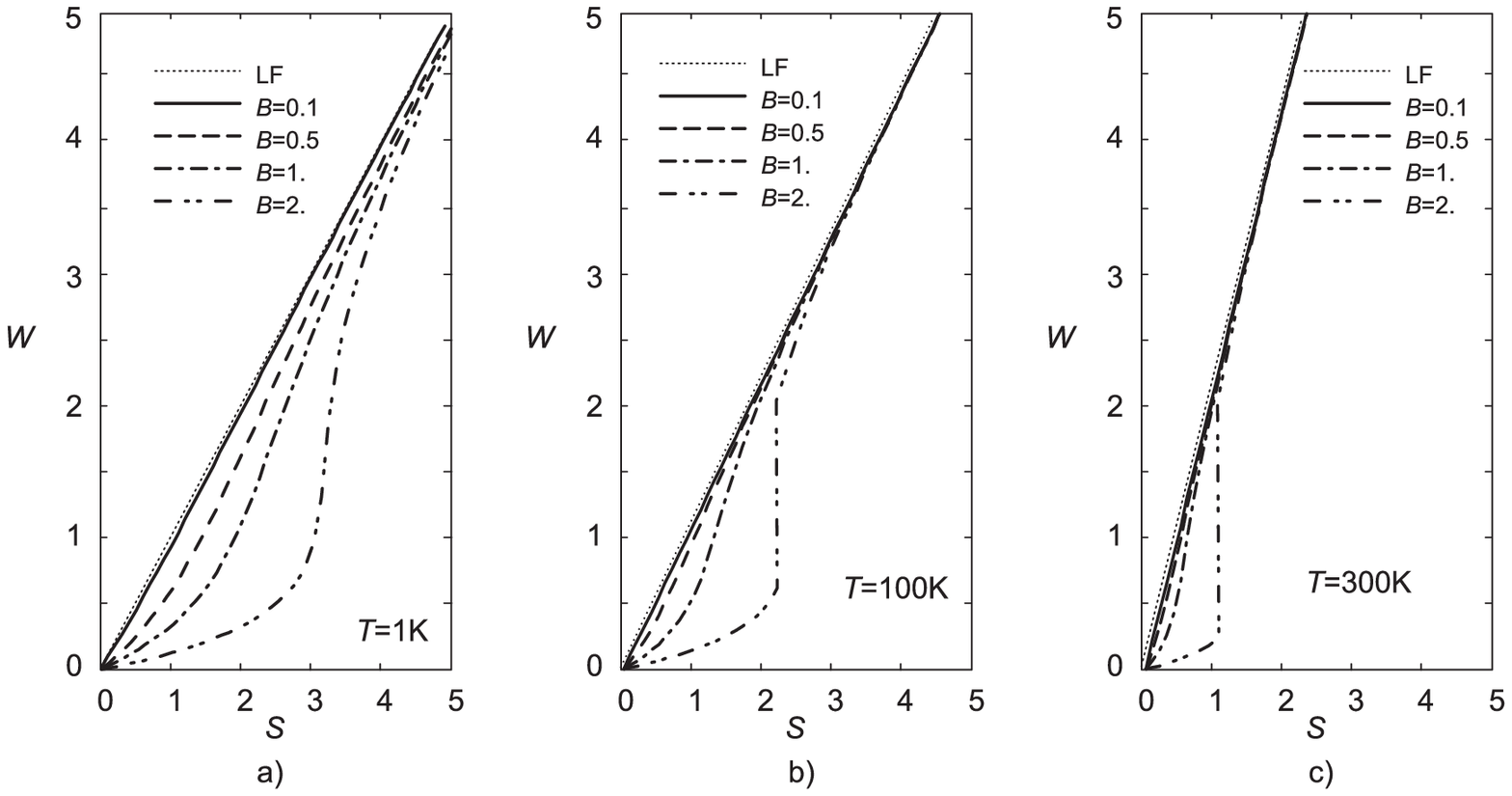}
  \includegraphics[height=.25\textheight]{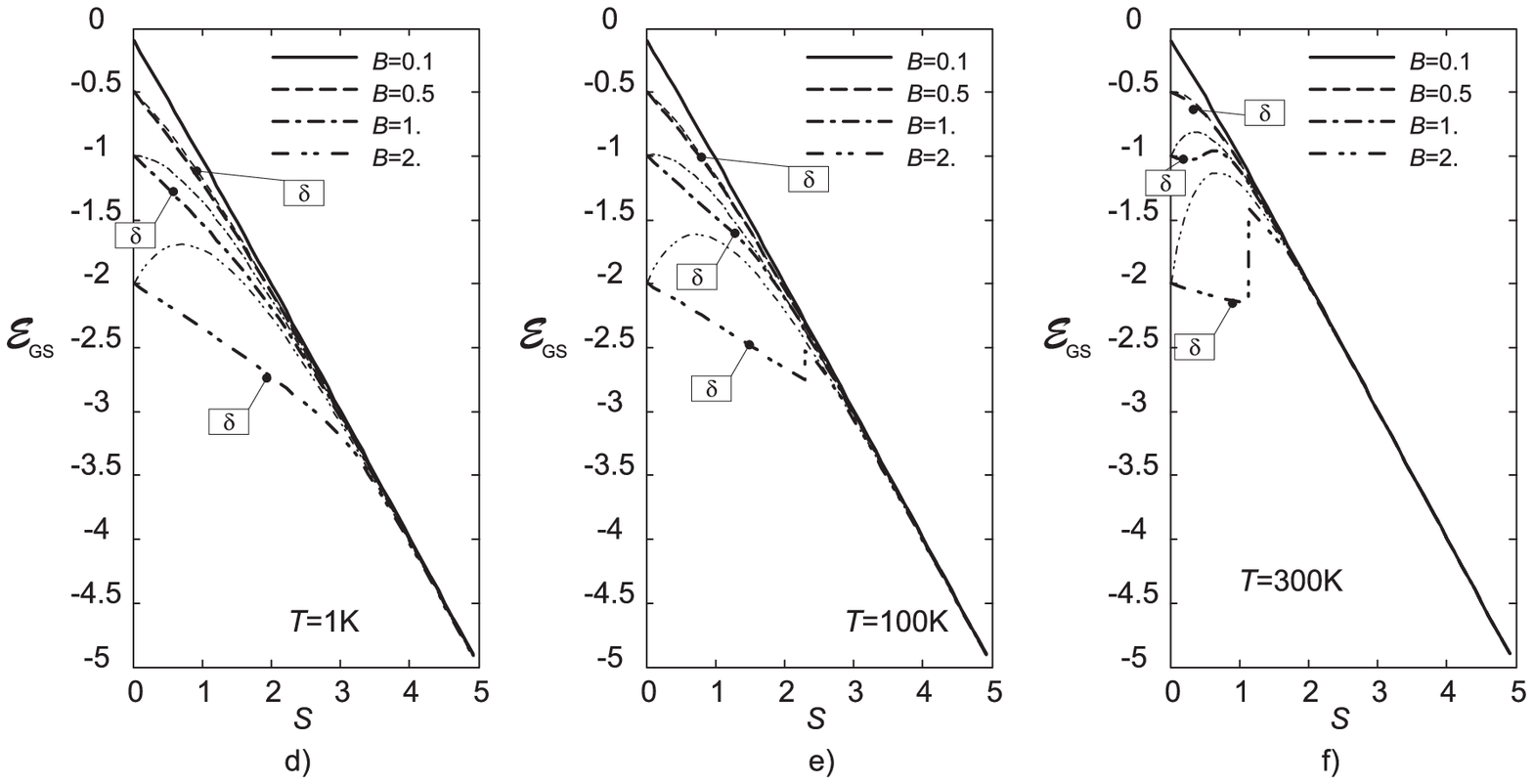}
  \caption{Dependence of the vibron dressing fraction $W$ (first row), and the vibron ground state energy $\mathcal{E}_{GS}$ (second row), on the coupling constant $S$, for various $B$ and $T$. The case of the vibron which moves along the covalent bonds and interacts with optical phonon modes.}\label{fig_2}
\end{center}
\end{figure}

\begin{figure}[H]
\begin{center}
  \includegraphics[height=.25\textheight]{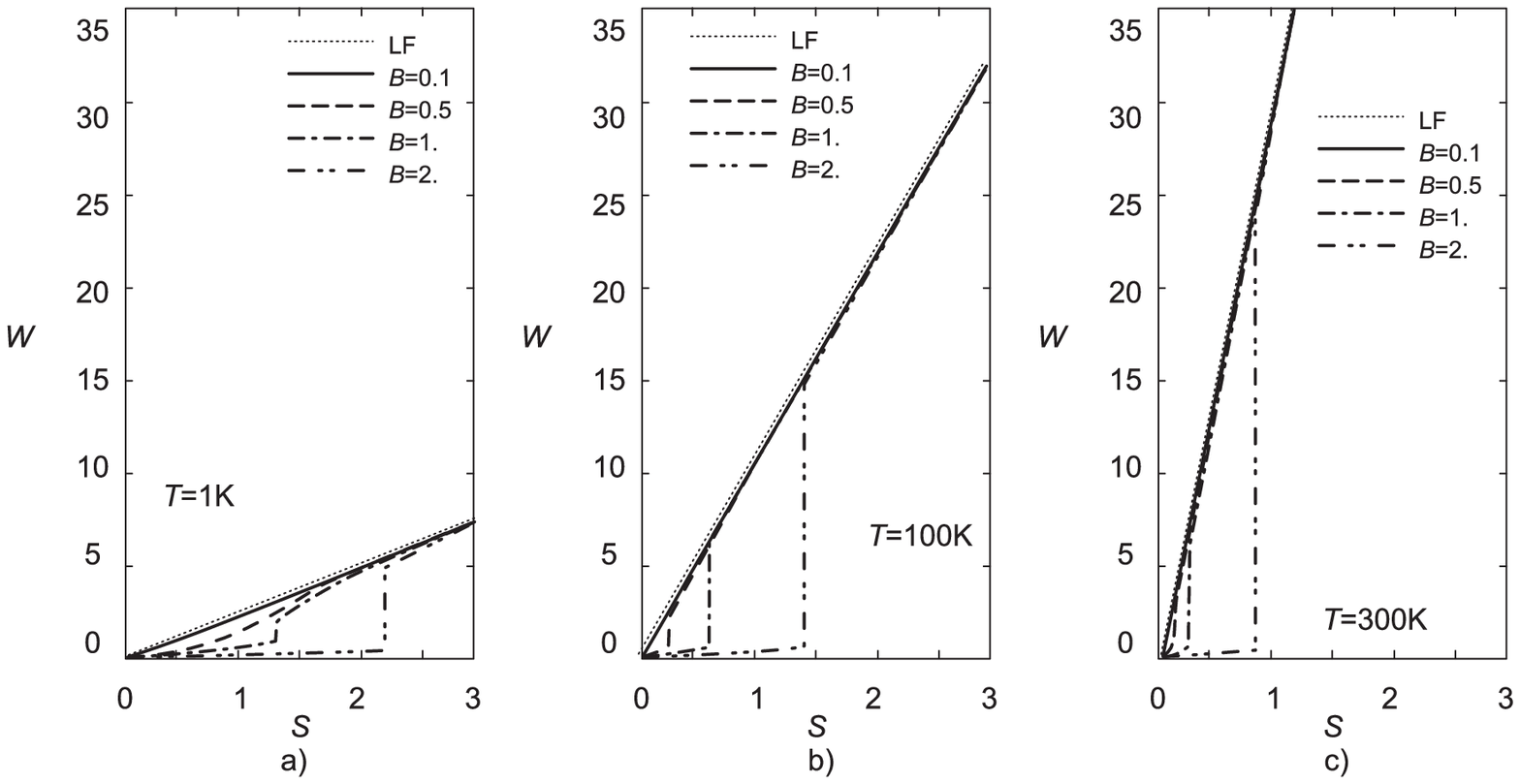}
  \includegraphics[height=.25\textheight]{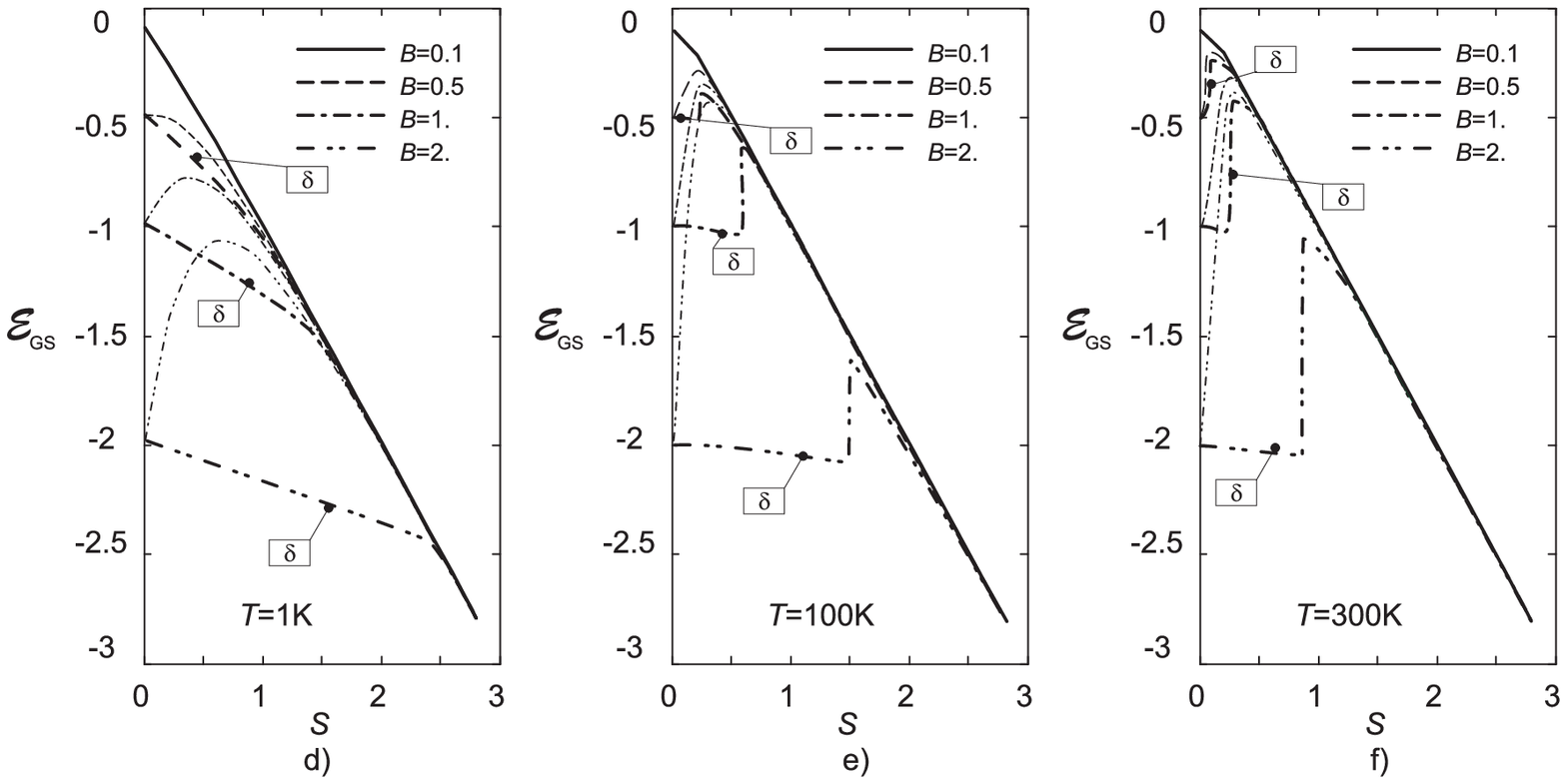}
  \caption{Dependence of the vibron dressing fraction $W$ (first row), and the vibron ground state energy $\mathcal{E}_{GS}$ (second row), on the coupling constant $S$, for various $B$ and $T$. The case of the vibron which moves along the hydrogen bonds and interact with acoustical phonon modes.}\label{fig_3}
\end{center}
\end{figure}

\begin{figure}[H]
\begin{center}
  \includegraphics[height=.25\textheight]{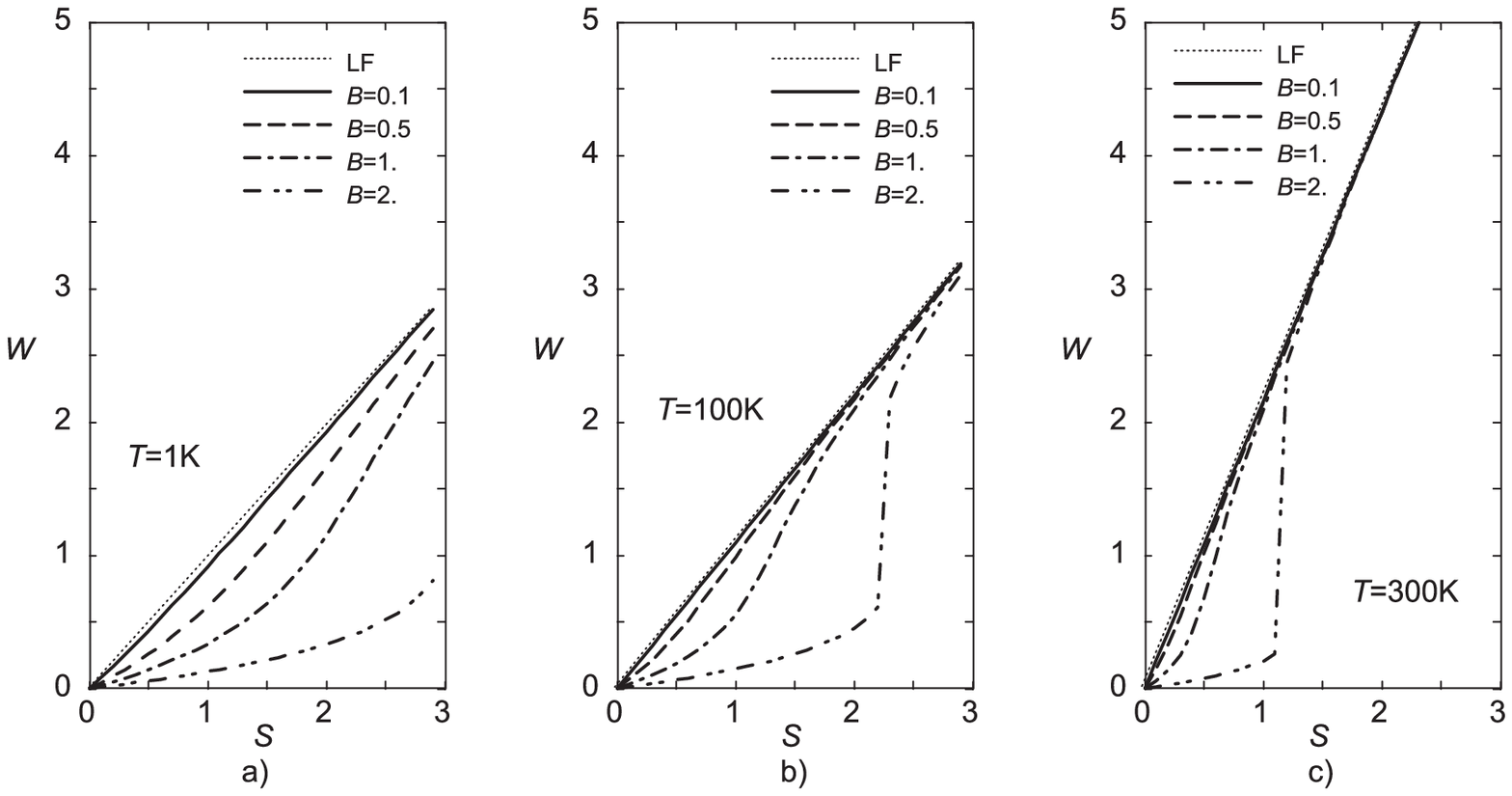}
  \includegraphics[height=.25\textheight]{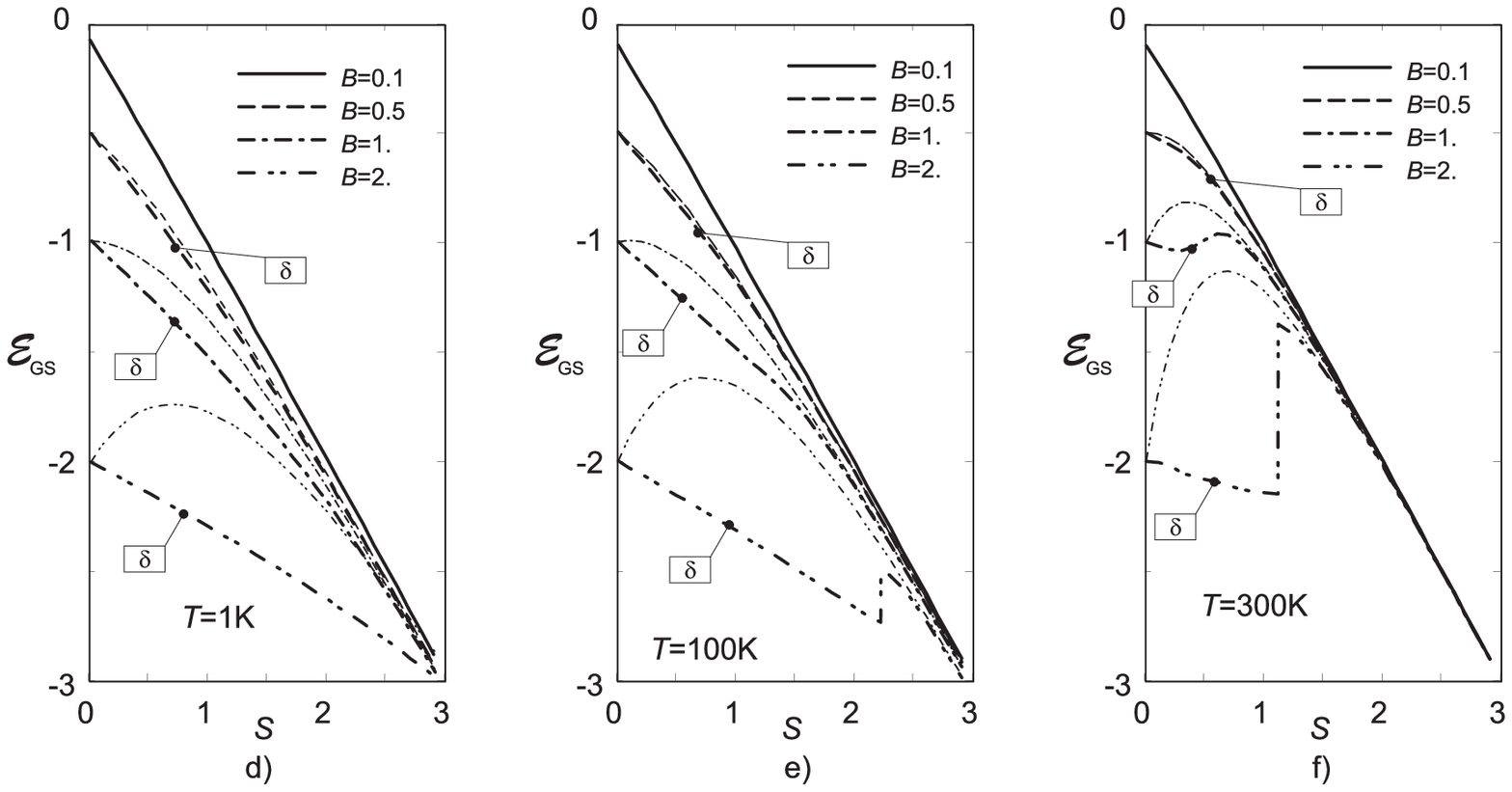}
  \caption{Dependence of the vibron dressing fraction $W$ (first row), and the vibron ground state energy $\mathcal{E}_{GS}$ (second row), on the coupling constant $S$, for various $B$ and $T$. The case of the vibron which moves along the hydrogen bonds and interacts with optical phonon modes.}\label{fig_4}
\end{center}
\end{figure}

One should be noted that from the above presented results  in the strong coupling limit and for all values of $B$ and $T$ the obtained values of both quantities are qualitatively very similar with ones derived by the usage of standard SP approach (i.e. with results obtained with using of the Lang--Firsov unitary transformation). This similarity increases with the decreasing of $B$. For those values of the system parameters both approaches predict heavy dressed vibron states. In addition,  with the growth of the system temperature and for fixed value of $B$, the values of the parameter $S$ for which our approach and standard SP theory yield similar results moves towards lower values. These observations seem to be not surprising, because of the framework of the standard small--polaron approach, the growth of the system temperature has practically the same effect as the growth of the coupling constant. This point appears by the consequence of the fact that the SP dressing fraction (and consequently quasiparticle effective mass) increases as temperature grows, but it also increases with the growth of coupling constant value \cite{IvicPB}.

On the other hand, for small and medium values of $S$ our results are significantly differs from ones predicted by standard LF theory. In contrast to the standard SP theory, in that limit our approach predicts slightly dressed or partially dressed vibron states (instead of characteristic parabolic shape, the adiabatic curves $\mathcal{E}_{GS}(S)$ has linear dependence of coupling constant). For fixed value of system temperature and $S$ this difference increase with increasing of $B$. Similarly, for fixed $S$ and $B$ the difference increases with increasing of $T$. All these conclusions are fulfilled for both types of the intermolecular forces between peptide units, and for both cases of the vibron--phonon interactions.

As it is demonstrated on Figs.1--4 there are two clearly distinguished areas in the system parameter space, where $W(S)$ and $\mathcal{E}_{GS}(S)$ display qualitatively different behaviors. In fact, for each $T$, type of the vibron--phonon interaction, and bond type there exists the critical adiabatic parameter $B_C$, such that, for all $B<B_C$, dressing fraction $W(S)$ and ground state energy $\mathcal{E}_{GS}(S)$ changes smoothly as a function of coupling constant $S$. In that case function $W(S)$ monotonically increases, while $\mathcal{E}_{GS}(S)$ for small values of $S$ slightly decreases, and after some value of $S$ it becomes very closely to linear decreasing function. Thus, for $B<B_C$ the vibron states for small values of $S$ corresponds to small--polaron band states with gradual transition towards the self--trapped ones as the case the coupling constant increases.

But when $B$ reaches some critical value $B_C$, adiabatic curves $W(S)$ have an infinity derivative at a point with abscissa $S_C$. In the same time, ground state energy $\mathcal{E}_{GS}(S)$ ceases to be smooth at the point with the same abscissa. The point in the system parameter space, with coordinates $(S_C,B_C)$ determines the "`critical point"'. By further increasing of the value of $B$ ($B>B_C$, i.e. for supercritical values of $B$), adiabatic curves are smooth up to a supercritical value of $S$, where they suffer the abrupt change. The main characteristic of the system in this region of parameter space is the spontaneous abrupt transition between the low dressed and high dressed states, which takes place through the discontinuous change in the magnitude of SP parameters, effective mass and ground state energy, in particular. As one can remark, with increasing of adiabatic parameter $B$, the $S_C$ moves to larger values. But, as system temperature $T$  increases, the $S_C$ moves to lower values (for fixed value of $B$). One can also remark that this observations are stronger expressed for vibron which delocalizes along the hydrogen bonds, especially in the case of the vibron interaction with acoustic phonon modes.

The existence of the critical and supercritical points is a consequence of the appearance of multiple local extremes of the function $\mathcal{F}_B(\delta)$.	In fact, in the case of supercritical values of $B$, and for small values of $S$, the function $\mathcal{F}_B(\delta)$ is increasing with respect to $\delta$. With increasing of $S$ it firstly appears one local minimum at a point which abscissa is $\delta \approx 0$. With further increasing of $S$, there appear two additional local extremes: one local maximum and one local minimum (new minimum appears at the point with abscissa $\delta\approx 1$). For smaller values of $S$, the minimum that corresponds to the lower value of $\delta$ is deeper, and consequently, for these values of $S$ and $B$ vibron is lightly dressed (small value of $\delta$ implies a small value of the dressing fraction). When $S$ reach $S_C$, both local minima have equal depths. This means that there exist two vibron states (light dressed vibron state and heavy dressed vibron state) which both correspond to the same minimal value of $\mathcal{F}_B$. With further increasing of $S$ the deeper minimum of $\mathcal{F}_B$ becomes the minimum that corresponds to heavy dressed vibron state ($\delta\approx1$). This model predicts a similar vibron nature for both types of intermolecular chains, and for both types of vibron--phonon interactions.

Above mentioned discontinuities are well known and sometimes have been considered as unphysical and the artifact of the variational methods mostly used in particular calculations. Nevertheless, in spite of all doubts the rapid increase of the polaron effective mass with the coupling constant in the adiabatic regime ($B\gg 1$) is the main characteristic of the so called SP crossover and represents the general conclusion regardless of the particular theoretical method which is used \cite{BIPRB40,ToyozawaPTP26,EminPRL,VenFish,Wagner,IadonisiPSSB203,IadonisiEL41}. Qualitatively precisely the same behavior has been undoubtedly confirmed recently by means of the numerically exact diagonalization \cite{AlvermannPRB81} of the Holstein model where  discontinuous transition\footnote{The notion of phase transition has been used in \cite{AlvermannPRB81} to emphasize its abrupt character.} from free electron to immobile small polaron $m_{eff}=\infty$ has been predicted. Moreover, the possible coexistence of free and localized states have been recently pointed out by Hamm and Tsironis \cite{HammPRB78} who had investigated the dimensionality effects on large to small polaron crossover in Holstein's model. The emergence of energy barrier separating the two types of solutions has been predicted irrespectively of the system dimensionality. This barrier is responsible for the double--peak structure of the wave function and its appearance has been interpreted as a clear signature of the coexistence of free (large) and localized (small) polaron solutions in a certain parameter range, which is in qualitative agreement with our results. Finally, it is interesting that there exists the close analogy between abrupt change from delocalized to localized vibron state in macromolecular chains  and thermally activated positronium state transitions in alkali halide cristals. It is known that the ground state of positronium in such structures is delocalized (Bloch-type), but there exists additional metastable state which is localized and it can be achieved by the increasing of the system temperature. The localization of the positronium was investigated experimentally \cite{HyodoJPSJ, KasaiJPSJ52,KasaiJPSJ57}, and theoretically \cite{BondarevPRB57,BondarevPRB58}, and it can be explained by the thermally activated transition from ground state to excited metastable state.

As we mentined above, the critical values of coupling constant $S_C$ and adiabatic parameter $B_C$ depends of the nature of the intermolecular bonds, phonon nature, as well as of the system temperature. This dependence is represented on Figs. 5 and 6 by dashed lines. Supercritical values of $S$ and $B$ are presented for a several values of $T$ by sets of "isothermal" curves (full lines).

\begin{figure}[H]
\begin{center}
  \includegraphics[height=.3\textheight]{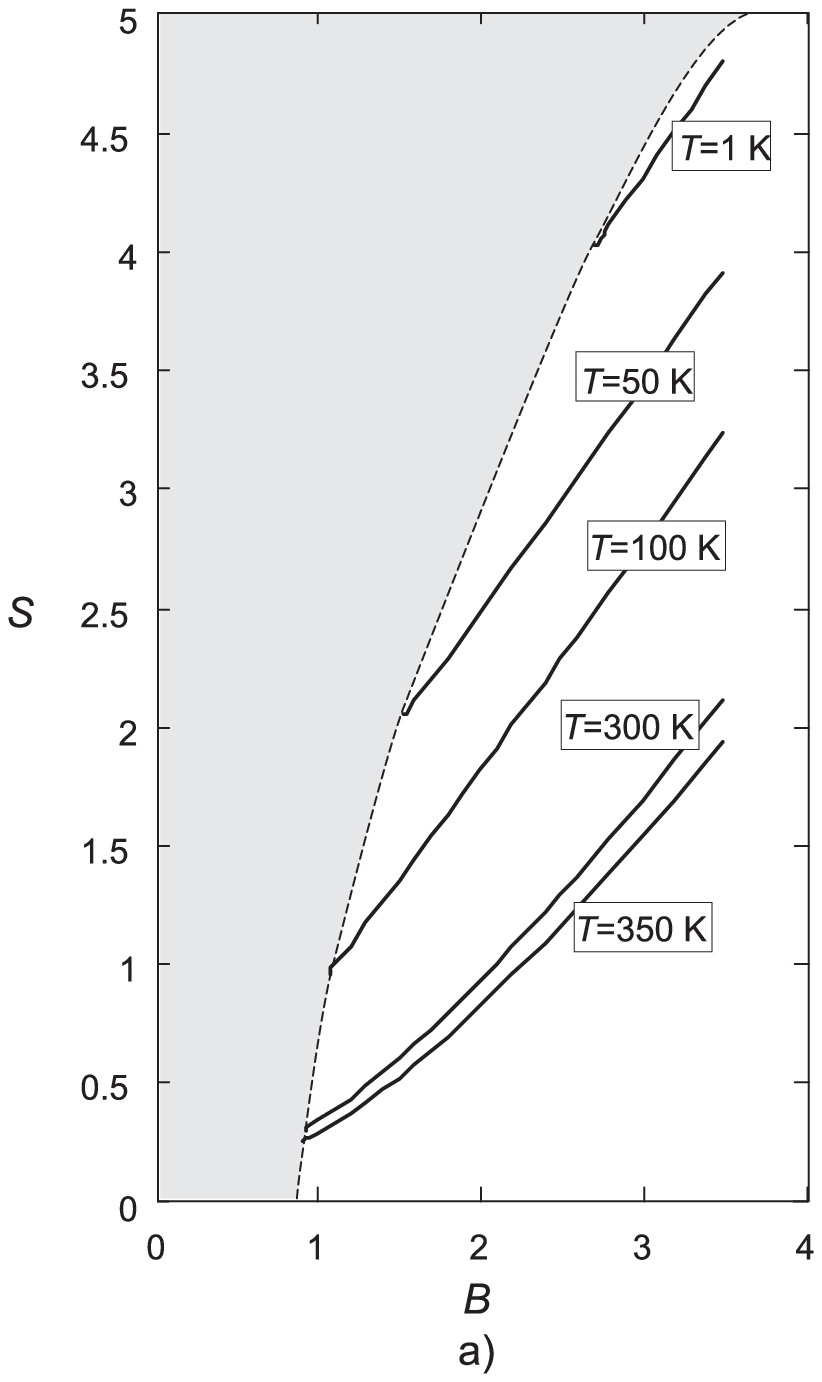}
  \includegraphics[height=.3\textheight]{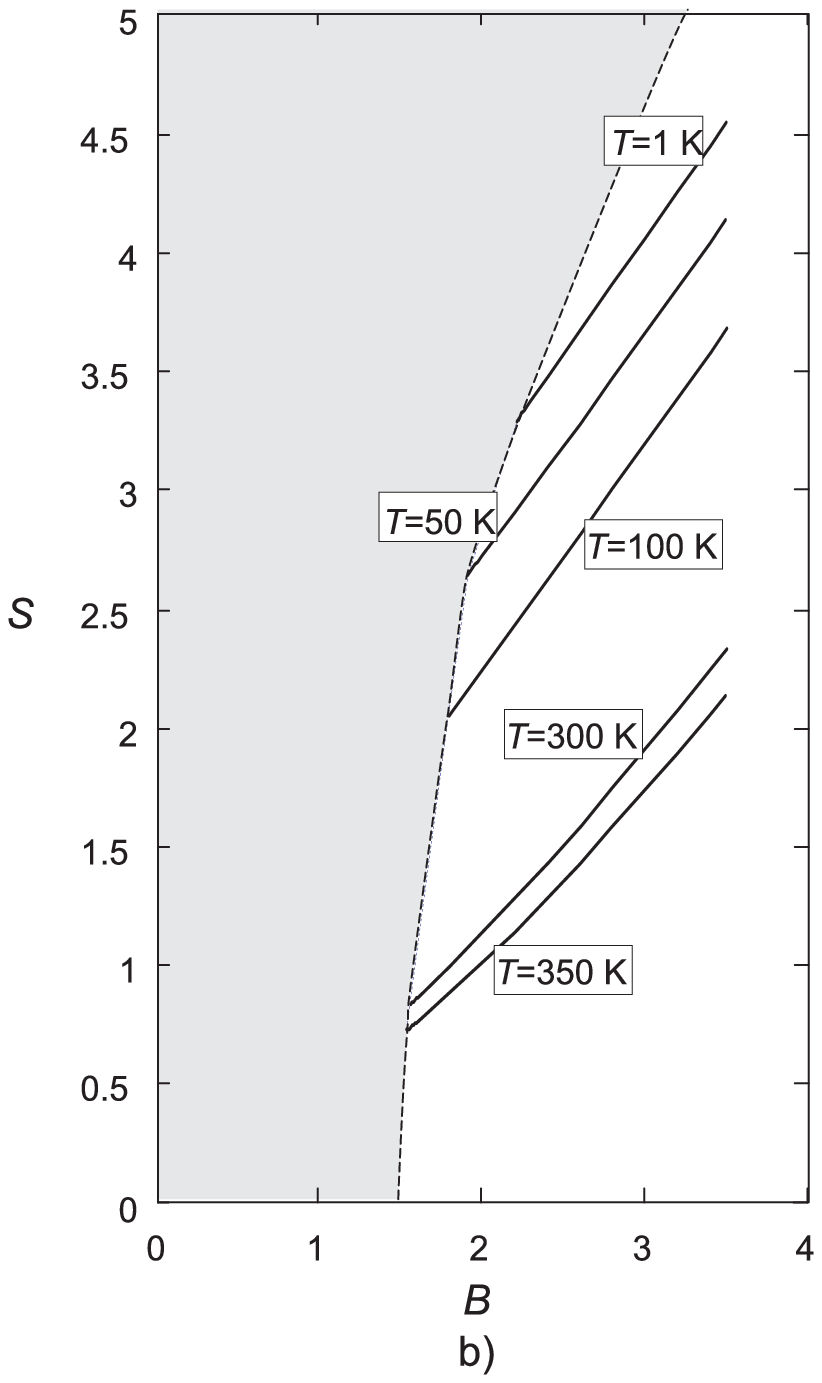}
  \caption{The dependence of the boundary values of the system parameters $(S,B)$ on temperature in the case of vibron delocalization along the covalent bonds: a) vibron interacting with acoustical phonon modes; b) vibron interacts with optical phonon modes.}\label{fig_5}
\end{center}
\end{figure}\begin{figure}[H]

\begin{center}
  \includegraphics[height=.3\textheight]{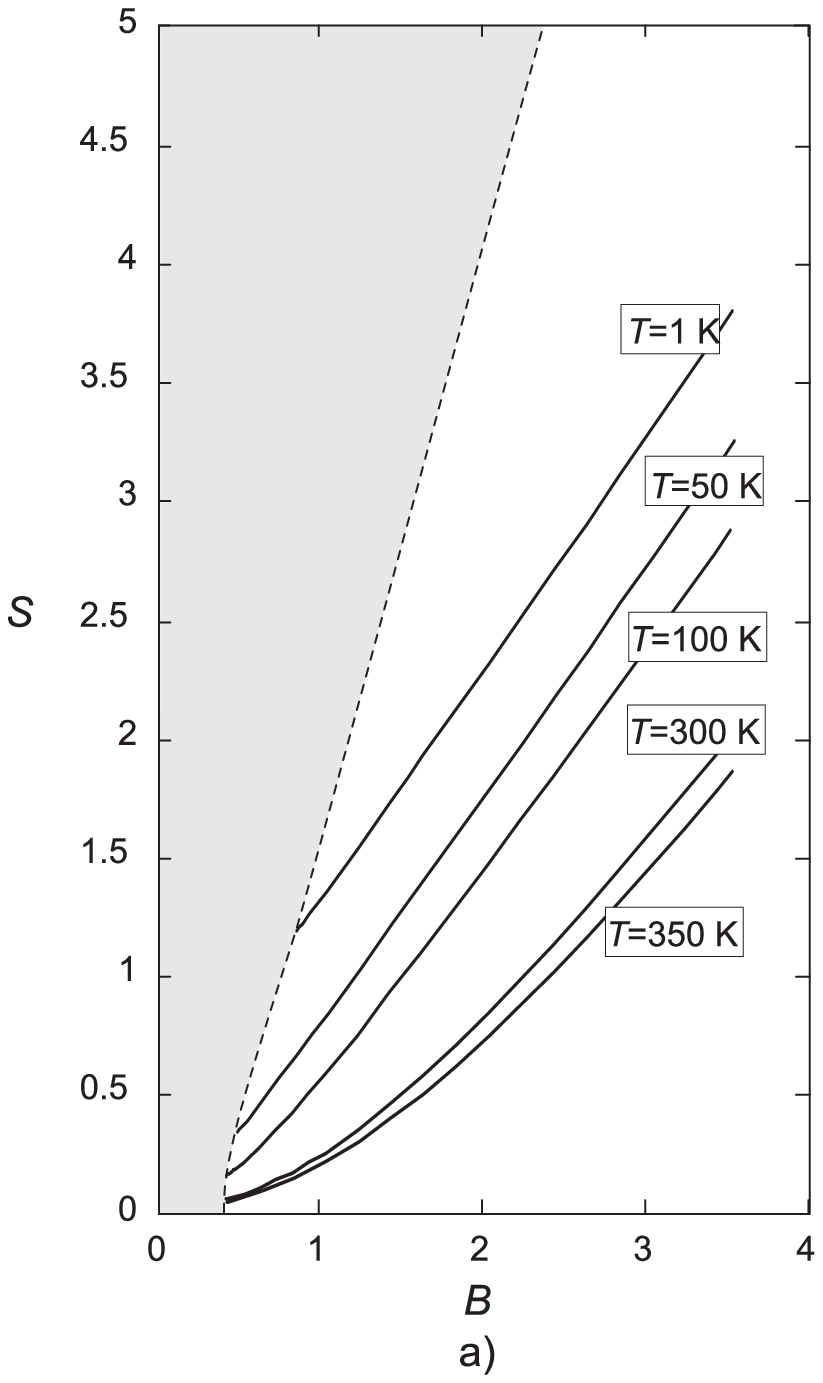}
  \includegraphics[height=.3\textheight]{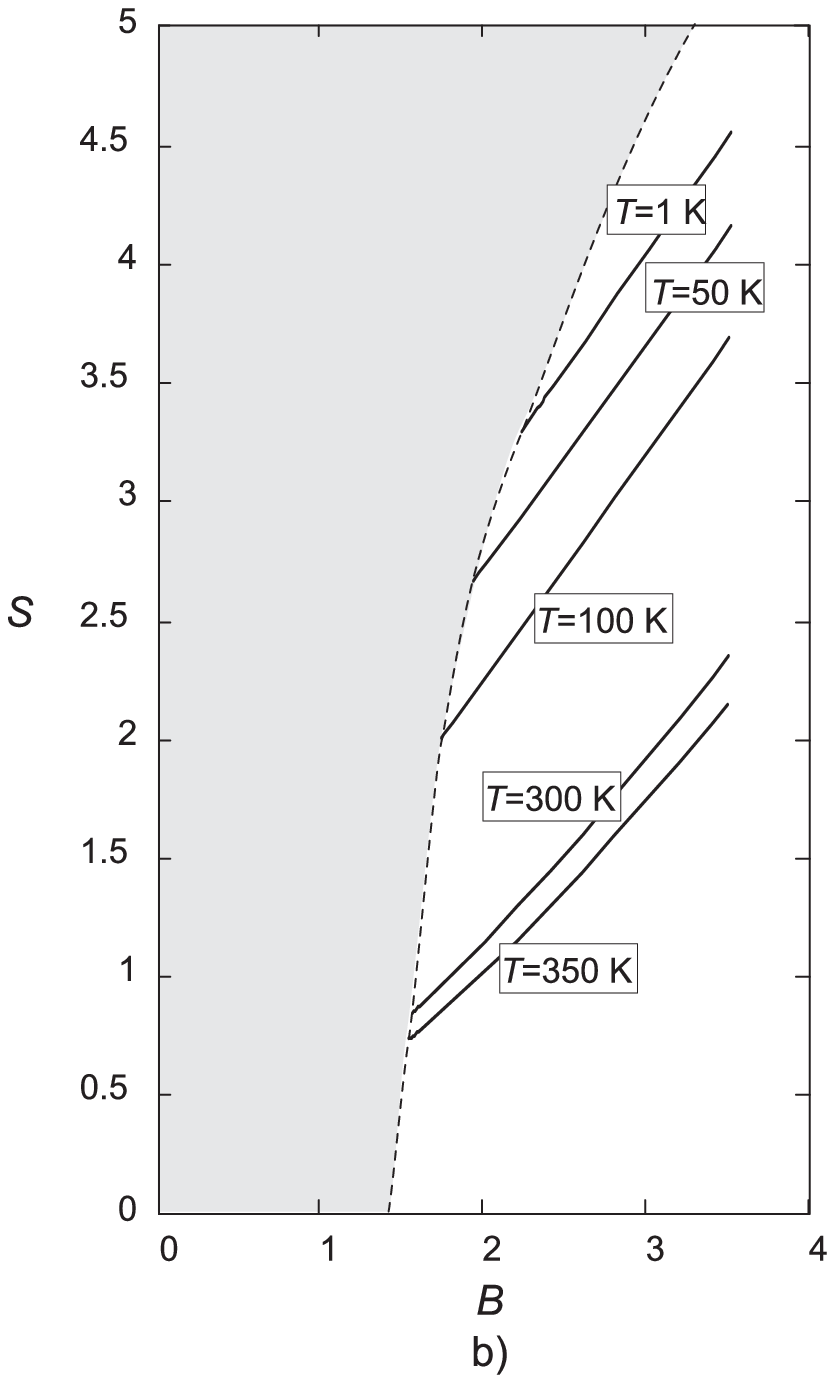}
  \caption{The dependence of the  boundary values of the system parameters $(S,B)$ on temperature in the case of vibron delocalization along the hydrogen bonds: a) vibron interacting with acoustical phonon modes; b) vibron interacts with optical phonon modes.}\label{fig_6}
\end{center}
\end{figure}

Dashed line divides the system parameter space on two areas: for the system parameter values that belong to the left area the curves $W(S)$ and $\mathcal{E}_{GS}(S)$ are smooth, and for larger values of $S$ they are very close to the ones predicted by the standard LF approach. Here, presented model predicts a significant difference with standard LF approach only for medium values of $S$, where obtained adiabatic curves  have a characteristic shape. The points that lie at right side from the dashed line are supercritical points, where $W(S)$ and $\mathcal{E}_{GS}(S)$ have an abrupt change.

The critical values of system parameters are shifted to lower values, with increasing of the system temperature. As an example, in the case of the vibron which delocalizes along the covalent bond and interacts with acoustical phonon modes, these values belong to the adiabatic and strong coupling region. They enters in non--adiabatic region approximately at $T=145$ K (at room temperatures $T=300 K$, we obtain $B_C=0.92$ and $S_C=0.303$. These values of the system parameters belong to non--adiabatic region of intermediate interaction. In the case when vibron interacts with optical phonon modes, $B_C$ belongs to adiabatic region for whole temperature range (at $T=300$ K, we obtain $B_C=1.55$ and $S_C=0.819$). However, in the case  of the vibron delocalization along the hydrogen bonds, and interaction with acoustical phonon modes, $B_C$ belongs to non--adiabatic region for all values of the system temperatures (at $T=1$ K we obtain $B_C=0.82$ and $S_C=1.179$), and at room temperatures it enters deeply in non--adiabatic region (at $T=300$ K we obtain $B_C=0.39$ and $S_C=0.056$)! In the case of vibron interaction with optical phonon modes obtained results are very similar to ones, obtained for vibron delocalization along the covalent bonds and its interaction with optical phonon modes.

\section{Conclusions}

Summarizing, the presented model of the $\alpha$--helicoidal macromolecular chain predicts that vibron self--trapping would results in the creation of partially dressed small--polaron band states. With the growth of the temperature, heavy dressed, practically localized small--polaron states may appear instead of the band states, irrespective on the nature of the bonds between two structural units. The values of the critical parameters for which the vibron nature changes are dependent on the system temperature and nature of the bonds. It is noticeable that values of the $B_C$ decrease with the rise of the system temperature.

In the case of an $\alpha$--helix structure our system model predict the following. When vibron interacts with optical phonon modes, the critical value of adiabatic parameter belongs to adiabatic region for all values of system temperature, as well as for both intermolecular bond types. In that case the critical value of coupling constant $S_C$ belongs to strong coupling limit at low temperatures, and reaches to intermediate coupling strength approximately at room temperatures.

When vibron interacts with acoustical phonon modes, $B_C$ enters into non--adiabatic region at $T=145$ K in the case of vibron delocalization along the covalent bonds. However, in the case of vibron delocalization along the hydrogen bonds, $B_C$ belongs to non--adiabatic region for all values of system temperatures.

Due to the fact that the values of the system parameters for the alpha--helix macromolecules belongs to the non--adiabatic and weak to intermediate coupling limit, it seems that at room temperatures in such systems may appear only heavy dressed vibron which moves along the covalent bonds, but for vibron which moves along the hydrogen bonds may appears weakly dressed vibron states.

Obtained results are similar with results obtained in the study of the electron polaron effective mass around the $\Gamma$ point of the Brillouin zone \cite{IadonisiPSSB203,IadonisiEL41}. Additionally, according to the study by Hamm and Tsironis \cite{HammPRB78}, it follows that the results obtained here are consistent, at least on a qualitative level, with the exact numeric ones.

In the latter respect, our study represents the background for investigations of the polaron properties in the three-dimensional structure of $\alpha$--helicoidal proteins as well as the boundary between coherent and incoherent polaron motion. According to some authors \cite{PouthierJCP123, ChristiansenPRE56} the phonon spectra in $\alpha$--helicoidal structures exhibits two acoustical phonon branches and one optical phonon branch. The first ones in long--wavelength limit are essentially polarized along the axis of the direction defined by the axis of the helix (longitudinal vibration mode) and torsional mode. The second branch refers to a radial motion of the helix structural elements (breathing mode). Due to the fact that the vibron--phonon coupling parameter for the three--dimensional problem depends of the projection of the phonon polarization vector on the direction of the vibron motion, it is necessary to investigate the vibron interaction with all three phonon modes in an appropriate manner. In addition, we stress the further studies require the inclusion of the mutual influence of both dressing fractions and evaluation of the temperature dependent mean lifetime of the localized states.


\end{CJK*}  
\end{document}